\documentclass[aps,prd,twocolumn,fleqn,superscriptaddress]{revtex4}

\usepackage{graphicx,color}
\usepackage{amsmath,amssymb,amsfonts}

\newcommand{\be}{\begin{equation}}
\newcommand{\ee}{\end{equation}}

\newcommand{\bea}{\begin{eqnarray}}
\newcommand{\eea}{\end{eqnarray}}

\def\be{\begin{equation}}
\def\ee{\end{equation}}
\def\beq{\begin{eqnarray}}
\def\eeq{\end{eqnarray}}

\def\({\left (}
\def\){\right )}
\def\[{\left [}
\def\[{\right ]}

\input amssym.def
\input amssym.tex

\begin{document}

\title{Thermalization of Strongly Coupled Field Theories}

\author{V.~Balasubramanian}
\affiliation{David Rittenhouse Laboratory, University of Pennsylvania, Philadelphia, PA 19104, USA}
\author{A.~Bernamonti}
\affiliation{Theoretische Natuurkunde, Vrije Universiteit Brussel, and
Int.~Solvay Inst., B-1050 Brussels, Belgium}
\author{J.~de~Boer}
\affiliation{Institute for Theoretical Physics, University of Amsterdam,
1090 GL Amsterdam, The Netherlands }
\author{N.~Copland}
\affiliation{Theoretische Natuurkunde, Vrije Universiteit Brussel, and
Int.~Solvay Inst., B-1050 Brussels, Belgium}
\author{B.~Craps}
\affiliation{Theoretische Natuurkunde, Vrije Universiteit Brussel, and
Int.~Solvay Inst., B-1050 Brussels, Belgium}
\author{E.~Keski-Vakkuri}
\affiliation{Helsinki Institute of Physics \& Dept.~of Physics,
FIN-00014 University of Helsinki}
\author{B.~M\"uller}
\affiliation{Department of Physics \& CTMS, Duke University, Durham, NC 27708, USA}
\author{A.~Sch\"afer}
\affiliation{Institut f\"{u}r Theoretische Physik, Universit\"{a}t Regensburg, D-93040 Regensburg, Germany}
\author{M.~Shigemori}
\affiliation{Kobayashi-Maskawa Institute for the Origin of Particles and the Universe,
Nagoya University, Nagoya 464-8602, Japan}
\author{W.~Staessens}
\affiliation{Theoretische Natuurkunde, Vrije Universiteit Brussel, and
Int.~Solvay Inst., B-1050 Brussels, Belgium}

\begin{abstract}
Using the holographic mapping to a gravity dual, we calculate 2-point functions, Wilson loops, and entanglement entropy in strongly coupled field theories in 2, 3, and 4 dimensions to probe the scale dependence of thermalization  following a sudden injection of energy. For homogeneous initial conditions, the entanglement entropy thermalizes slowest, and sets a timescale for equilibration that saturates a causality bound.   The growth rate of entanglement entropy density is nearly volume-independent for small volumes, but slows for larger volumes. In this setting, the UV thermalizes first.
\end{abstract}

\maketitle

It is widely believed that the observed nearly inviscid hydrodynamics  of relativistic heavy ion collisions at collider energies is an indication that the matter produced in these nuclear reactions is strongly coupled \cite{Gyulassy:2004zy}. Some such strongly coupled field theories can be studied using the holographic duality between gravitational theories in asymptotically anti-de Sitter (AdS) spacetimes and quantum field theories on the boundary of AdS.   The thermal state of the field theory is represented by a black brane in AdS, and near-equilibrium dynamics  is studied in terms of perturbations of the black hole metric.  A key remaining challenge is to understand the far from equilibrium process of thermalization. The AdS/CFT correspondence relates the approach to thermal equilibrium in the boundary theory to black hole formation in the bulk. 

Recent works studied the gravitational collapse of energy injected into AdS$_5$ and the formation of an event horizon \cite{Chesler:2008hg}. These works started from locally anisotropic metric perturbations near the AdS boundary and studied the rate at which isotropic pressure was established by examining the evolution of the stress tensor.  By studying gravitational collapse induced by a small scalar perturbation, the authors of \cite{Bhattacharyya:2009uu} concluded that local observables  behaved as if the system thermalized almost instantaneously.  Here we model the equilibrating field configuration in AdS by an infalling homogeneous thin mass shell \cite{earlycollapse,Lin:2008rw} and study how the rate of thermalization varies with spatial scale and dimension.  We consider 2d, 3d and 4d field theories dual to gravity in asymptotically AdS$_3$, AdS$_4$ and AdS$_5$ space-times, respectively.  Our treatment of 2d field theories is analytic. 

Expectation values of local gauge-invariant operators, including the energy-momentum tensor and its derivatives, provide valuable information about the applicability of viscous hydrodynamics but  cannot be used to explore the scale dependence of deviations from thermal equilibrium.  Equivalently,   in the dual gravitational description these quantities are only sensitive to the metric close to the AdS boundary.  Nonlocal operators, such as Wilson loops and 2-point correlators of gauge-invariant operators, probe the thermal nature of the quantum state on extended spatial scales.  In the AdS language, these probes reach deeper into the bulk space-time, which corresponds to probing further into the infrared of the field theory. They are also relevant to the physics probed in relativistic heavy ion collisions, e.g.\ through the jet quenching parameter $\hat{q}$ \cite{Kovner:2001vi} and the color screening length.

A  global probe of thermalization is the entanglement entropy $S_A$ \cite{Calabrese:2009qy,Nishioka:2009un} of a domain $A$, measured after subtraction of its vacuum value.   In the strong coupling limit, it has been proposed that $S_A$ for a region $A$ with boundary $\partial A$ in the field theory is proportional to the area of the minimal surface $\gamma$ in AdS whose boundary coincides with $\partial A$: $S_A = {\rm Area}(\gamma)/4G_N$, where $G_N$ is Newton's constant \cite{Nishioka:2009un}. Thus, for a $(d=2)$-dimensional field theory,  $S_A$ is the length of a geodesic curve in AdS$_3$ (studied in \cite{AbajoArrastia:2010yt}); for $d=3$, $S_A$ is the area of a 2d sheet in AdS$_4$ (studied in \cite{Albash:2010mv}); and for $d=4$, $S_A$ is the volume of a 3d region in AdS$_5$.   In $d=3$ the exponential of the area  of the minimal surface that measures $S_A$ also computes the expectation value of the Wilson loop that bounds the minimal surface.   Wilson loops in $d=4$ correspond to 2d minimal surfaces as well. 

First, we consider equal-time 2-point correlators of gauge invariant operators ${\cal O}$ of large conformal dimension $\Delta$.  In the dual supergravity theory this correlator can be expressed, in the semiclassical limit, in terms of the length ${\cal L}(\mathbf{x},t)$ of the bulk geodesic curve that connects the endpoints 
on the boundary: 
$
\langle{\cal O}(\mathbf{x},t){\cal O}(0,t)\rangle \sim \exp[ - \Delta\, {\cal L}(\mathbf{x},t) ] 
$ \cite{Balasubramanian:1999zv}.
When multiple such geodesics exist, one has to consider steepest descent contours to determine the contribution from each geodesic.

\begin{figure}[t]
\centering
\centering
\includegraphics[width=0.3\linewidth]{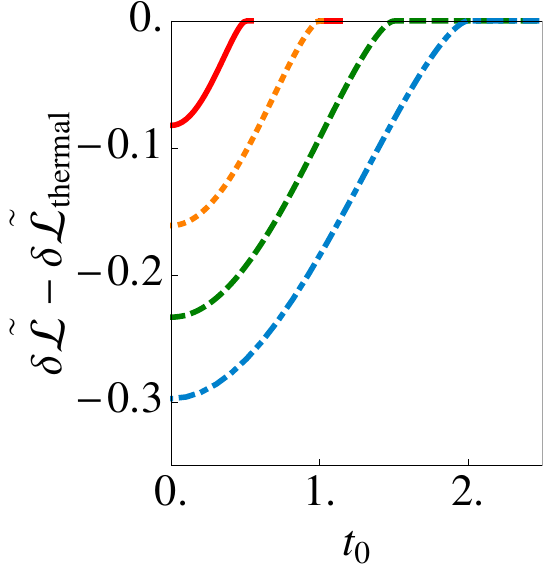}
\includegraphics[width=0.3\linewidth]{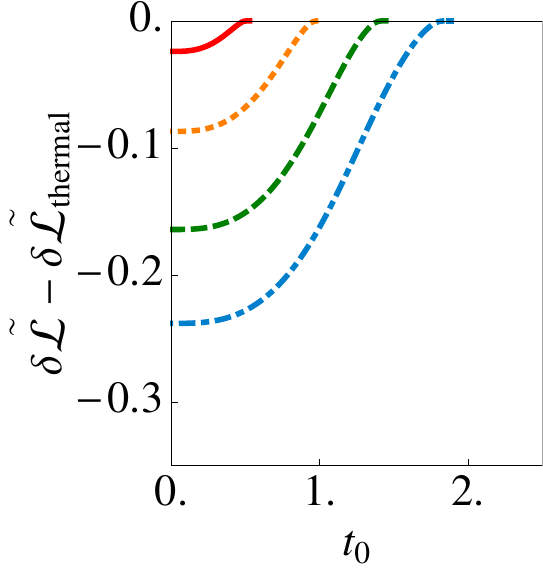}
\includegraphics[width=0.3\linewidth]{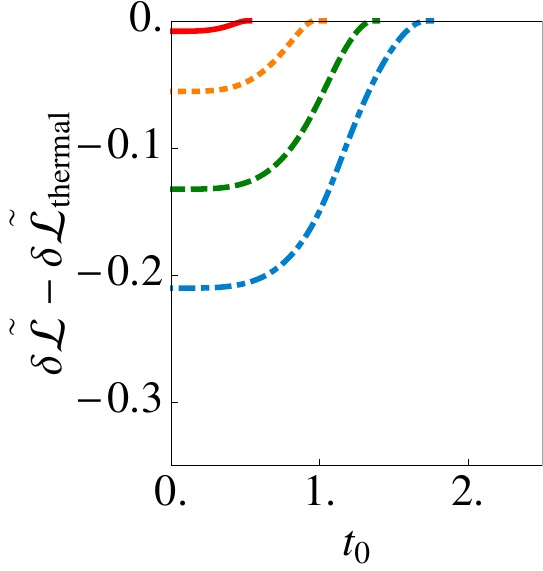}
\caption{$\delta \tilde {\cal L} - \delta \tilde {\cal L}_{{\rm thermal}}$ ($\tilde{\cal L} \equiv {\cal L} /  \ell$) as a function of  boundary time $t_0$ for $d=2,3,4$ (left,right, middle) for a thin shell ($v_0 = 0.01$).   The boundary separations are $\ell = 1,2,3,4$ (top to bottom curve).  All quantities are given in units of $M$.  These numerical results match analytical results for $d=2$ as $v_0 \to 0$.}
\label{fig:deltaL}
\end{figure}

We consider a $d+1$-dimensional infalling shell geometry described in Poincar\'e coordinates by the Vaidya metric
\be
ds^2 = \frac{1}{z^2}\left[-\left(1 - m(v)z^d\right) dv^2 - 2 dz\, dv + d\mathbf{x}^2 \right] ,
\label{eq:Vaidya}
\ee
where $v$ labels  ingoing null trajectories, and we set the AdS radius to 1. The boundary is at $z=0$, where  $v$ coincides with the observer time $t$. The mass function of the infalling shell is 
\be
m(v)=(M/2)\,\left( 1+\tanh(v/v_0) \right) ,
\label{eq:mv}
\ee
where $v_0$ determines the  thickness of a shell falling along $v =0$. The metric interpolates between vacuum AdS inside the shell and an AdS black brane geometry with Hawking temperature $T=dM^{1/d}/4\pi$ outside the shell. 2-point functions agree with those of a boundary field theory at thermal equilibrium only if they are dominated by geodesics that stay outside the shell. 

The geodesic length ${\cal L}$ diverges due to contributions near the AdS boundary. We introduce an ultraviolet cut-off $z_0$ and define a renormalized correlator  $\delta {\cal L}={\cal L}+2 \ln(z_0/2)$ by removing the divergent part of the correlator in the vacuum state (pure AdS). The  renormalized equal-time 2-point function is
$
\langle{\cal O}(\mathbf{x},t){\cal O}(0,t)\rangle_{\rm ren} \sim \exp[ - \Delta\, \delta {\cal L}(\mathbf{x},t) ] .
$
We compute the renormalized correlator as a function of $\mathbf{x}$ and $t$ in a state evolving towards thermal equilibrium and compare it to the corresponding thermal correlator. In the bulk, this amounts to computing geodesic lengths in a collapsing shell geometry and comparing them to geodesic lengths in the black brane geometry ($\delta{\cal L}_{{\rm thermal}}$) resulting from the collapse.

We study geodesics with boundary separation $\ell$ in the $x$ direction in AdS$_3$, AdS$_4$ and AdS$_5$ modified by the infalling shell. The  endpoint locations are denoted as $(v,z,x)=(t_0,z_0,\pm\ell/2)$, where $z_0$ is the UV cut-off.  The lowest point of the geodesic in the bulk is the midpoint located at $(v,z,x)=(v_*,z_*,0)$. Geodesics are obtained by solving  differential equations for the functions $v(x)$ and $z(x)$ with these boundary conditions and are unique in the infalling shell background.  The length of the geodesics is 
$
{\cal L}(\ell,t_0) = 2 \int_{0}^{\ell/2} dx\, z_*\, z(x)^{-2} .
$
In empty AdS, this gives the renormalized geodesic length $\delta{\cal L}_{{\rm AdS}} = 2 \ln(\ell / 2)$.

\begin{figure}[t]
\centering
\includegraphics[width=0.27\linewidth]{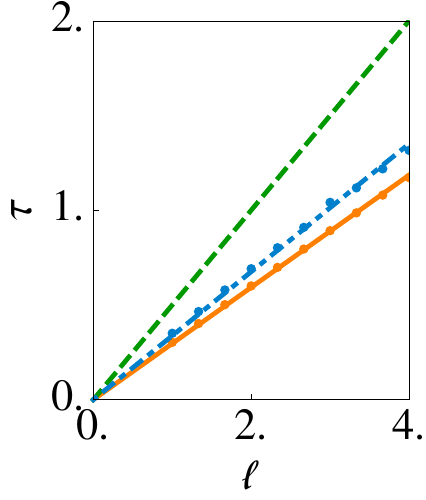}
\includegraphics[width=0.27\linewidth]{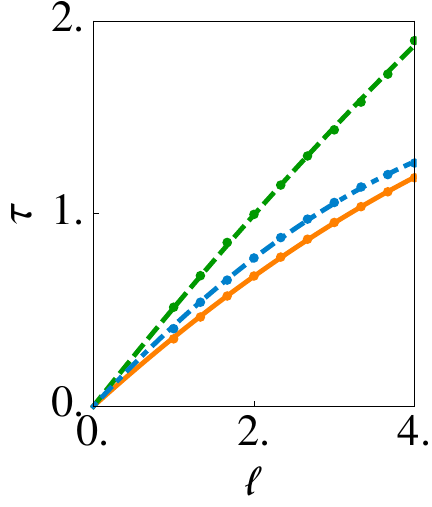}
\includegraphics[width=0.27\linewidth]{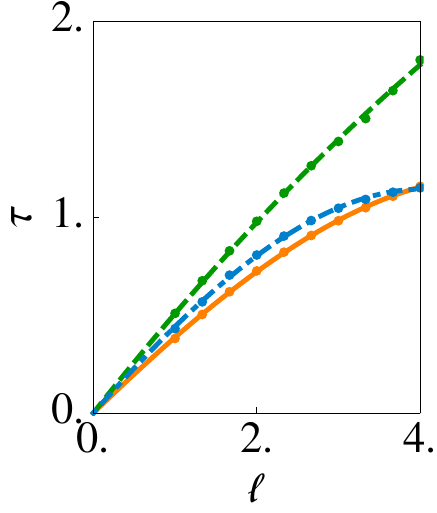}
\caption{Thermalization times ($\tau_{{\rm dur}}$,  top line; $\tau_{{\rm max}}$, middle line; $\tau_{1/2}$, bottom line) as a function of spatial scale for $d=2$ (left), $d=3$ (middle) and $d=4$ (right) for a thin shell ($v_0 = 0.01$).   All thermalization time scales are linear in $\ell$ for $d=2$, and deviate from linearity for $d=3,4$.  
}
\label{fig:thermtime}
\end{figure}

A numerical solution for the length of geodesics crossing the shell in the $d=2$ (AdS$_3$) case was obtained in \cite{AbajoArrastia:2010yt}.  We checked that physical  results do not depend significantly on the shell thickness when $v_0$ is small, and then derived an analytical solution in the $v_0\to 0$ limit:
\be\delta{\cal L}(\ell,t_0 ) = 2 \ln \left[ \frac{\sinh(\sqrt{M} t_0)}{\sqrt{M}  s(\ell,t_0)}  \right]  \, ,
\label{ads3shellgeod}
\ee
where $s(\ell,t_0) \in [0,1]$ is parametrically defined by:
\bea
\ell  &=& \frac {1}{\sqrt M} \left[ \frac{2 c}{s\rho} 
  + \ln \left( \frac{2(1+c)\rho^2+2s\rho-c}{2(1+c)\rho^2-2s\rho-c} \right) \right]  \, ,
\nonumber \\
2\rho  &=&  \coth(\sqrt{M} t_0) + \sqrt{\coth^2(\sqrt{M} t_0) - \frac{2c}{c+1}} \, ,
\eea
with $c = \sqrt{1 - s^2}$ and $\rho = (\sqrt{M}z_{{\rm c}})^{-1}$.  Here $z_{{\rm c}}$ is the radial location of the intersection between the geodesic and the shell.   For any given $\ell$,  at sufficiently late times, the geodesic lies entirely in the black brane background outside the shell. In this case the length is 
\be
\delta {\cal L}_{\rm thermal}(\ell) = 2 \ln \left[ (1/\sqrt M) \sinh \left( \sqrt{M} \ell /2 \right) \right]\, ,
\label{BTZgeod}
\ee
representing the result for thermal equilibrium.

We use these analytic relations in $d=2$ and find $\delta {\cal L}(\ell,t_0)$ in $d=3,4$ by numerical integration.  We measure the approach to thermal equilibrium by comparing $\delta{\cal L}$ at any given time with the late time thermal result (see Fig.~\ref{fig:deltaL}).  In any dimension, this compares the logarithm of the 2-point correlator at different spatial scales with the logarithm of the thermal correlator.   For $d=2$, the same quantity measures by how much the entanglement entropy at a given spatial scale differs from the entropy at thermal equilibrium. 

Various thermalization times can be extracted from Fig.~\ref{fig:deltaL}. For any spatial scale we can ask for: (a) the time $\tau_{{\rm dur}}$ until full thermalization (measured as the time when the geodesic between two boundary point just grazes the infalling shell), (b) the half-thermalization time $\tau_{1/2}$, which measures the duration for the curves to reach half of their equilibrium value, (c) the time $\tau_{{\rm max}}$ at which thermalization proceeds most rapidly, namely the time for which the curves in Fig.~1 are steepest.  These are plotted in Fig.~\ref{fig:thermtime}. 
 In $d=2$ we can analytically derive the linear relation $\tau_{{\rm dur}} \equiv \ell/ 2$, as also observed in  \cite{AbajoArrastia:2010yt}.

\begin{figure}[t]
\centering
\includegraphics[width=0.31\linewidth]{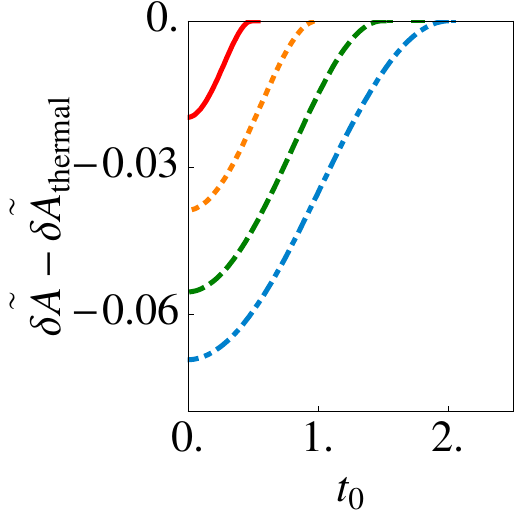}
\includegraphics[width=0.3\linewidth]{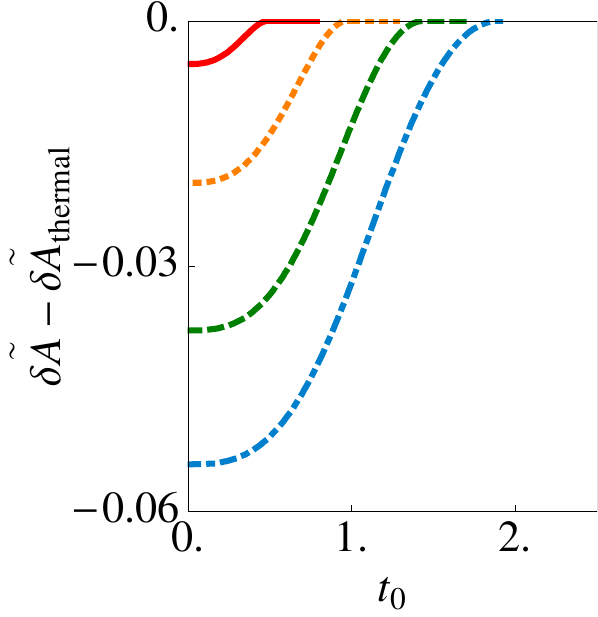}
\includegraphics[width=0.3\linewidth]{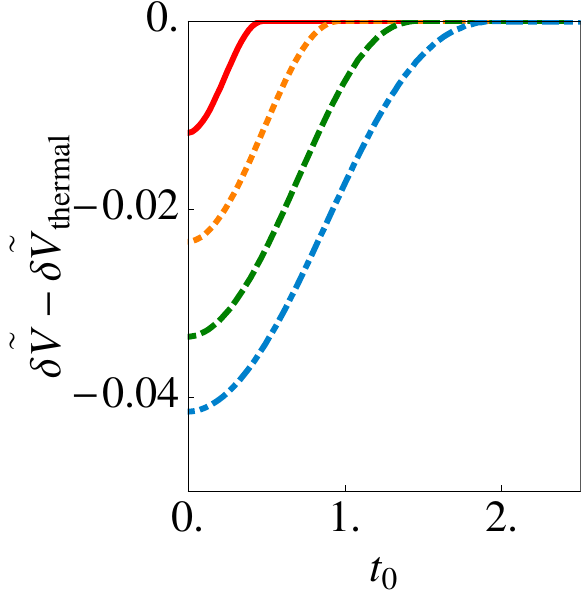}
\caption{$\delta \tilde{\cal A} - \delta \tilde{\cal A}_{{\rm thermal}}$ ($\tilde{\cal A} \equiv {\cal A}/\pi R^2$; left and middle panels)  and $\delta \tilde{ V} - \delta \tilde{ V}_{{\rm thermal}}$ ($\tilde{V} \equiv V/(4 \, \pi R^3 / 3)$; right panel)  as a function of  $t_0$ for radii $R = 0.5, 1, 1.5, 2$ (top curve to bottom curve) and mass shell parameters $v_0 = 0.01$, $M=1$, in $d=3$ (left panel) and $d=4$ (middle and right panel) field theories.
} 
\label{DL-Avt-2}
\end{figure} 

The linearity of $\tau_{{\rm dur}}(\ell)$ in 2d is expected from general arguments in conformal field theory \cite{Calabrese:2009qy}, and the coefficient is as small as possible under the constraints of causality. The thermalization time scales $\tau_{1/2}$ and $\tau_{{\rm max}}$ for 3d and 4d field theories (Fig.~\ref{fig:thermtime}, middle and right) are sublinear in the spatial scale.  In the range we study, the complete thermalization time $\tau_{{\rm dur}}$ deviates slightly from linearity, and is somewhat shorter than $\ell/2$.  We will later discuss whether a rigorous causality bound for thermalization processes exists or not.    

In 2d ``quantum quenches'' where a pure state prepared as the ground state of a Hamiltonian with a mass gap is followed as it evolves according to a different, critical Hamiltonian, a nonanalytic feature was found where thermalization at a spatial scale $\ell$ is completed abruptly at $\tau_{{\rm dur}}(\ell)$ \cite{Calabrese:2009qy,AbajoArrastia:2010yt}.  An analogous feature is evident in Fig.~\ref{fig:deltaL} (left) as a sudden change in the slope at $\tau_{{\rm dur}}$, smoothed out only by the small non-zero thickness of the shell, or equivalently, by the intrinsic duration of the injection of energy.    We find a similar (higher-order) non-analyticity for $d=3,4$ (Fig.~\ref{fig:deltaL}, middle and right) and expect this to be a general consequence of abrupt injection of energy in any dimension.

Fig.~\ref{fig:thermtime} shows that complete thermalization of the equal-time correlator is first observed at short length scales, or large momentum scales (see also \cite{Lin:2008rw}). While this behavior follows directly in our setup with a shell falling in from the (``UV'') boundary of AdS, this ``top-down'' thermalization contrasts with the behavior of weakly coupled gauge theories even with energy injected in the UV. In the ``bottom-up'' scenario \cite{Baier:2000sb} applicable to that case, hard quanta of the gauge field do not equilibrate directly by randomizing their momenta, but  gradually degrade their energy by radiating soft quanta, which fill up the thermal phase space and equilibrate by collisions among themselves. This bottom-up scenario is  linked to the infrared divergence of the splitting functions of gauge bosons and fermions in perturbative gauge theory. It contrasts with the ``democratic'' splitting properties of excitations in strongly coupled SYM theory that favor an approximately equal sharing of energy and momentum \cite{Hatta:2008tx}.

\begin{figure}[t]
\centering
\includegraphics[width=0.27\linewidth]{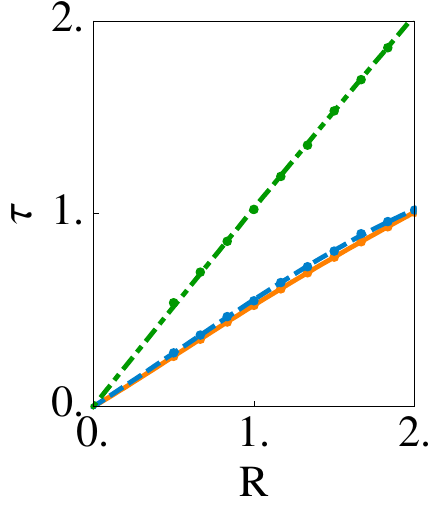}
\includegraphics[width=0.27\linewidth]{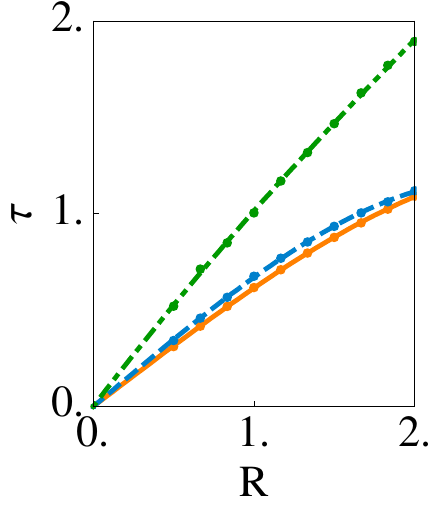}
\includegraphics[width=0.27\linewidth]{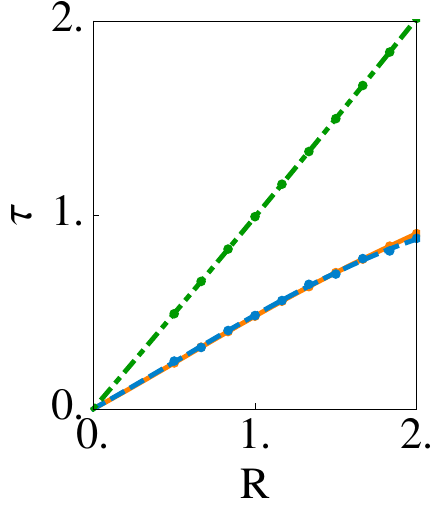}
\caption{Thermalization times ($\tau_{{\rm dur}}$,  top line; $\tau_{{\rm max}}$, middle line; $\tau_{1/2}$, bottom line) as a function of the diameter for circular Wilson loops in $d=3,4$ (left, middle) and for entanglement entropy of spherical regions in $d=4$ (right). 
}
\label{DL-Avt-3}
\end{figure}

The thermal limit of the Wightman function that we studied above is a necessary but not a sufficient condition for complete thermalization.  To examine whether thermalization proceeds similarly for other probes, we also studied entanglement entropy and spacelike Wilson loop expectation values in 3d (following \cite{Albash:2010mv}) and 4d field theories.  Entanglement entropy in 3d field theories is holographically related to minimal surfaces in AdS$_4$  and hence to the logarithm of the expectation value of Wilson loops. We considered circular loops of radius $R$ in $d=3,4$. The minimal spacelike surface in AdS$_{d+1}$ whose boundary is this circular loop extends into the bulk space radially and into the past.  The tip occurs at $(v_*,z_*,{\bf x}={\bf 0})$.   The cross section at fixed $z$ and $v$ is a circle, and thus the surface is parameterized in terms of the radii $\rho$ of these circles. The overall shape minimizes the  action for the two functions $z(\rho)$ and $v(\rho)$:
\be
{\cal A}[R] = 2\pi \int_{0}^{R} d\rho \frac{\rho}{z^2}\sqrt{1- \left(1-m(v)z^d\right)v'^2 - 2 z' v'} ,
\label{eq:loopaction}
\ee
where $z'(\rho)=dz/d\rho$ etc.  The resulting Euler-Lagrange equations can be numerically integrated.  
We regularize the area by subtracting the divergent piece of the area in ``empty'' AdS: $\delta{\cal A}[R] = {\cal A}[R] - \left( R/z_0 \right)$.  Entanglement entropy of spherical volumes in $d=4$ is similarly computed in terms of minimal volumes in AdS$_5$ by minimizing an equation similar to (\ref{eq:loopaction}) and defining $\delta V[R]$ by subtracting the divergent volume in ``empty'' AdS.

The deficit area $\delta{\cal A} -\delta{\cal A}_{\rm thermal}$ for Wilson loops in $d=3,4$ and the deficit volume $\delta V-\delta V_{\rm thermal}$ are plotted in Fig.~3 for several boundary radii $R$ as a function of the boundary time $t_0$. By subtracting the thermal values, we can observe the deviation from equilibrium for each spatial scale at a time $t_0$. Comparing the three thermalization times defined earlier as a function of the loop diameter (Fig.~4), we find that for the entanglement entropy in $d=3,4$, the complete thermalization time $\tau_{{\rm dur}}(R)$ is close to being a straight line with unit slope over the range of scales that we study (as observed in \cite{Albash:2010mv} for $d=3$). On the other hand, for Wilson loops in $d=4$, $\tau_{{\rm dur}}(R)$  deviates somewhat from linearity and is shorter than $R$.

Our thermalization times for Wilson loop averages and entanglement entropy seem remarkably similar to those for 2-point correlators (after noting that $R$ here is the radius of the thermalizing region and $\ell$ in Fig.~2 is the diameter).   Slightly ``faster-than-causal'' thermalization, possibly due to the homogeneity of the initial configuration, seems to occur for the probes that do not correspond to entanglement entropy in each dimension.  For the latter, the thermalization time is linear in the spatial scale and saturates the causality bound. As the actual thermalization rate of a system is set by the slowest observable, our results suggest that in strongly coupled theories with a gravity dual, thermalization occurs ``as fast as possible'' at each scale, subject to the constraint of causality. Taking the thermal scale $\ell \sim \hbar/T$ as length scale, this suggests that for strongly coupled matter $\tau_{{\rm dur}} \sim 0.5\hbar/T$, in particular  $\tau_{{\rm dur}} \sim 0.3$ fm/$c$ at heavy ion collider energies ($T\approx 300-400$ MeV),  comfortably short enough to account for the experimental observations.

\begin{figure}[t]
\centering
\includegraphics[width=1.1in,height=1.1in]{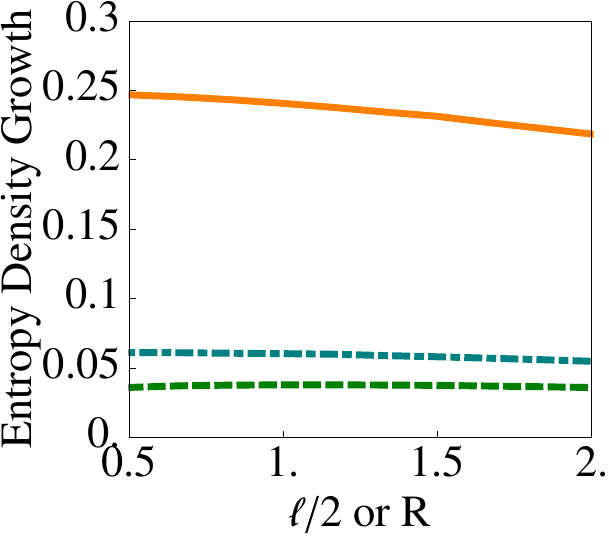}
\includegraphics[width=1.1in,height=1.1in]{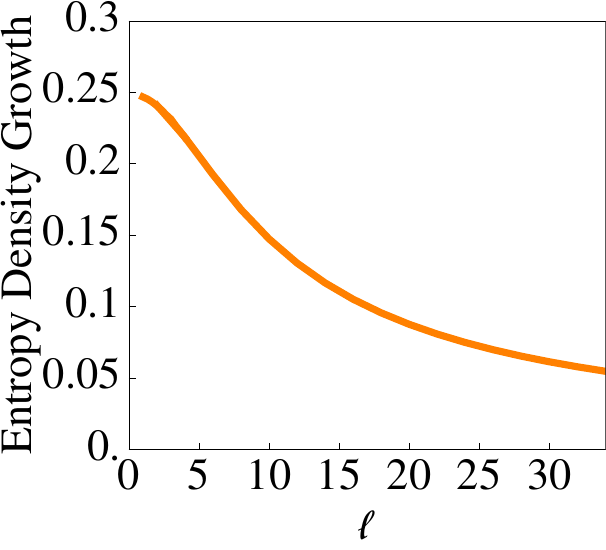}
\includegraphics[width=1.1in,height=1.1in]{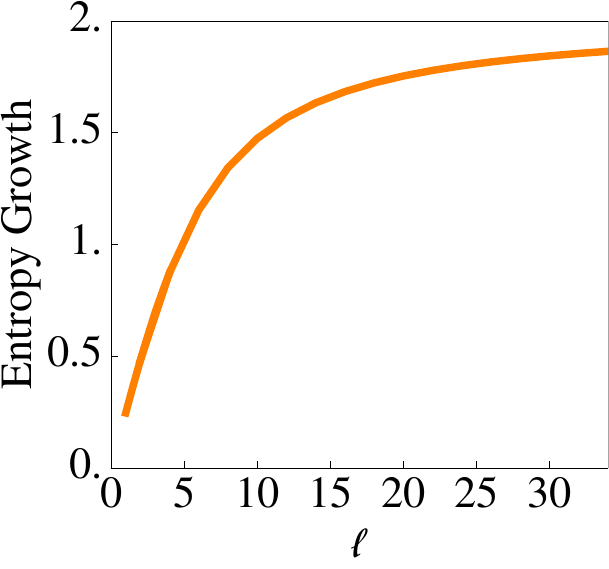}
\caption{(Left) Maximal growth rate of entanglement entropy {\it density} vs. radius of entangled region for $d=2,3,4$ (top to bottom).  (Middle) Same plot for $d=2$, larger range of $\ell$.  (Right) Maximal entropy growth rate for $d=2$.}
\label{fig:KSAdS3}
\vspace{-0.1in}
\end{figure}

The average growth rate of the {\it coarse grained} entropy in nonlinear dynamical systems is measured by the Kolmogorov-Sina\"i (KS) entropy rate $h_{\rm KS}$ \cite{Sinai}, which is given by the sum of all positive Lyapunov exponents.  For a classical SU(2) lattice gauge theory in 4d $h_{\rm KS}$  has been shown to be proportional to the volume \cite{Bolte:1999th}. For a system starting far from equilibrium, the KS entropy rate generally describes the rate of growth of the coarse grained entropy during a period of linear growth after an initial dephasing period and before the close approach to equilibrium \cite{Latora:1999xx}.   Here we observe similar linear growth of {\it entanglement} entropy density in $d=2,3,4$ (Figs.~\ref{fig:deltaL}a, \ref{DL-Avt-2}a, \ref{DL-Avt-2}c).  For small boundary volumes, the growth rate of entropy {\it density} is nearly independent of the boundary volume (almost parallel slopes in Figs.~\ref{fig:deltaL}a, \ref{DL-Avt-2}a, \ref{DL-Avt-2}c and nearly  constant maximal growth rate in Fig.~\ref{fig:KSAdS3}a). Equivalently, the growth rate of the entropy is proportional to the volume -- suggesting that entropy growth is a local phenomenon.    However, in $d=2$ where our analytic results enable study of large boundary volumes $\ell$, we find that the growth rate of the entanglement entropy density changes for large $\ell$, falling asymptotically as $1/\ell$ (Fig.~\ref{fig:KSAdS3}b).  Equivalently, the entropy has a growth rate that approaches a constant limiting value for large $\ell$ (Fig.~\ref{fig:KSAdS3}c), and thus cannot arise from a local phenomenon.  This behavior suggests that entanglement entropy and coarse grained entropy have different dynamical properties. 


We have investigated the scale dependence of thermalization following a sudden injection of energy in 2d, 3d, and 4d strongly coupled field theories with gravity duals. The entanglement entropy sets a time scale for equilibration that saturates a causality bound. The relationship between the entanglement entropy growth rate and the KS entropy growth rate defined by coarse graining of the phase space distribution raises interesting questions.

{\em Acknowledgments:} We thank V.~Hubeny for helpful discussions and E.~Lopez for comments on an earlier version of the manuscript. This research is supported  by the Belgian Federal Science Policy Office, by FWO-Vlaanderen, by the Foundation of Fundamental Research on Matter (FOM), by the DOE, by the BMBF, and by Academy of Finland. AB and WS are Aspirant FWO.


\end{document}